\documentclass[onecolumn,notitlepage,12pt]{article}
\usepackage{amsfonts}
\usepackage{amsmath}
\usepackage{graphicx}
\usepackage{graphicx}

\input{tcilatex}

\begin{document}

\title{A complex adaptive systems approach to the kinetic folding of RNA}
\author{Wilfred Ndifon \\
Departments of Biology and Mathematics,\\
Morgan State University, Baltimore, MD 21251\\
windi1@mymail.morgan.edu}
\date{July 20, 2005}
\maketitle

\begin{abstract}
The kinetic folding of RNA sequences into secondary structures is modeled as
a complex adaptive system, the components of which are possible RNA
structural rearrangements (SRs) and their associated bases and base pairs.
RNA bases and base pairs engage in local stacking interactions that
determine the probabilities (or fitnesses) of possible SRs. Meanwhile,
selection operates at the level of SRs; an autonomous stochastic process
periodically (i.e., from one time step to another) selects a subset of
possible SRs for realization based on the fitnesses of the SRs. Using
examples based on selected natural and synthetic RNAs, the model is shown to
qualitatively reproduce characteristic (nonlinear) RNA folding dynamics such
as the attainment by RNAs of alternative stable states. Possible
applications of the model to the analysis of properties of fitness
landscapes, and of the RNA sequence to structure mapping are discussed.
\end{abstract}

\baselineskip18pt

\section{Introduction}

RNA takes part in a variety of important cellular activities, including
protein synthesis, intron splicing, gene silencing, and genome rearrangement
(Lee et al., 2002; Mochizuki et al., 2002; Gratias \& Betermier, 2003; Yang
et al., 2003). Considering the extensive functional repertoire of RNA
molecules, it is of interest to determine their (functional) native
structures and to understand the (kinetic) process by which they fold into
such structures. The native structures of RNA molecules can be computed
efficiently, at the functionally relevant secondary structure level, using
free-energy minimization methods (Hofacker et al., 1994; Mathews et al.,
2004) or, in cases where sufficient homologous sequences are available, by
phylogenetic comparisons (Woese \& Pace, 1993; Cannone et al., 2002).

Several algorithms have been developed for studying the kinetic process by
which RNA molecules fold into their native secondary (Mironov \& Kister,
1985; Morgan \& Higgs, 1996; Flamm et al., 2000; Zhang \& Chen, 2000;
Wolfinger et al., 2003; Tang et al., 2004; Ndifon \& Nkwanta, 2005) and
tertiary (Abrahams et al., 1990; Gultyaev et al., 1990; Isambert, 2000;
Xayaphoumine et al., 2003) structures. The majority of these algorithms
(e.g., see Isambert, 2000; Xayaphoumine et al., 2003; Ndifon \& Nkwanta,
2005) operate on a helix-based move-set, involving the formation and
dissociation of entire RNA helices. On the other hand, a few of the
algorithms (e.g., see Zhang \& Chen, 2000; Wolfinger et al., 2003) operate
on a pair-based move-set; they model the RNA folding process as a
time-series of structural transitions, involving the formation and
dissociation of individual RNA base pairs.

Flamm and colleagues (Flamm, 1998; Flamm et al., 2000) have extended the
afore-mentioned pair-based move-set by introducing the concept of base pair 
\textit{shifting,} which makes possible description of the biological
process of defect-diffusion, believed to be an important feature of the 
\textit{in vivo} folding kinetics of RNA (Poerschke, 1974a). The folding
model developed in this paper implements this extended pair-based move-set
and is inspired by the theory of complex adaptive systems (see Section 3).
The applicability of the model is illustrated through several examples based
on selected natural and synthetic RNAs (see Section 4). In particular, the
folding kinetics of the yeast tRNA$^{Phe}$ is shown to be strongly
influenced by modifications to specific hairpin loops. In addition, a
characteristic optimal folding temperature $T_{opt}$ $\left( \approx
313K\right) $ of tRNA$^{Phe}$, at which the native state exhibits maximal
accessibility, is identified. Furthermore, estimates are obtained for the
population dynamics of two alternative stable states of SV11, an RNA species
that is replicated by $Q\beta $ replicase (Zamora et al., 1995).

The remainder of this paper is organized as follows. In Section 2, we
present some concepts related to RNA secondary structures and folding
kinetics. We introduce the theory of complex adaptive systems and discuss
details of the new folding model in Section \ref{themodel}. In Section \ref%
{applics}, we apply the model to some example problems and discuss other
possible applications in Section \ref{concl}.

\section{Background information}

\subsection{RNA secondary structure}

Let $X$ be an arbitrary RNA sequence of length $n$. We think of $X$ as a
string $X=x_{1}x_{2}\cdots x_{n}$ defined over the nucleotide alphabet $%
\left\{ A,C,G,U\right\} $. The nucleotides or bases of $X$ have a propensity
to \textit{pair} (or form canonical and non-canonical bonds) with each
other. A pair formed by the bases $x_{i}$ and $x_{j}$, $i<j$, is denoted by $%
\left( i,j\right) $. Two base pairs $\left( i,j\right) $ and $\left( i\prime
,j\prime \right) $ are said to be compatible if either $i<i\prime <j\prime
<j $ or $i<j<i\prime <j\prime $. If we let $H$ be the set of possible pairs
that can be formed by the bases of $X$, then a secondary structure $S$ of $X$
can be thought of as a set of mutually compatible base pairs drawn from $H.$
The multiset consisting of all subsets of $H$, including the empty set
(i.e., the open chain), forms the conformation space of $X$, denoted here by 
$\zeta (X)$. Note that incompatible base pairs form pseudoknots, which are
prohibited from occurring in the folding model developed in this paper. The
model can, however, be readily extended to allow the formation of
pseudoknots once reliable thermodynamic parameters for such tertiary
structural elements become available.

\subsection{Kinetic folding}

The kinetic folding of an RNA sequence $X$ at the coarse-grained secondary
structure level can be thought of as a time-series of structural
transitions, mediated by a set of operations called the move-set (Flamm,
1998). Each operation or \textit{move} converts one secondary structure $%
S_{i}\in \zeta \left( X\right) $ into another $S_{j}\in \zeta \left(
X\right) $. For each structure $S_{i}\in \zeta \left( X\right) $, the
move-set defines a neighborhood $N\left( S_{i}\right) $ such that $S_{j}\in $
$\zeta \left( X\right) $ belongs to $N\left( S_{i}\right) $ if and only if $%
d\left( S_{i},S_{j}\right) \leq d\in $ $%
\mathbb{R}
^{+}$, where $d\left( S_{i},S_{j}\right) $ is the (Hamming) distance between
structures $S_{i}$ and $S_{j}$ and $d$ is the "move distance". Only \textit{%
moves} that convert $S_{i}$ into some $S_{j}\in N\left( S_{i}\right) $ are
legal. The probability of a legal move is given by a rate equation, an
example of which is the Metropolis rule (Metropolis et al., 1953):

\begin{equation}
k_{ij}=\left\{ 
\begin{array}{c}
e^{\frac{-\left( G_{j}-G_{i}\right) }{RT}}\text{, if }G_{j}<G_{i} \\ 
1\text{, \ \ \ \ \ \ \ \ \ \ \ if }G_{j}\geq G_{i}%
\end{array}%
\right.
\end{equation}%
where $P\left( S_{i}\rightarrow S_{j}\right) $ is the probability of
converting $S_{i}$ into $S_{j}$ $\in N\left( S_{i}\right) $ by a single 
\textit{move}; $G_{i}$ and $G_{j}$ are, respectively, the free energies of $%
S_{i}$ and $S_{j}$, computed using a suitable choice of free-energy
parameters (e.g., Mathews et al., 1999).

Conventional Monte Carlo RNA folding algorithms execute in each time step $%
t+\delta t$ a \textit{move} that converts the nascent RNA structure $S_{t}$
into some structure $S_{t+\delta t}$, where $S_{t}$ denotes the secondary
structure of $X$ at time $t$. The time-ordered series of structures $\left\{
S_{t}\right\} _{t\geq 0},$ with $S_{t}\in \zeta \left( X\right) $ and $%
S_{t=0}$ the open chain, is called a folding trajectory of $X$. The folding
time $\tau _{f}$ associated with a given folding trajectory is the minimum
value of $t$ for which $S_{t}$ is the native structure of $X$. A folding
trajectory satisfying the following condition is called a folding path
(Flamm et al., 2000): $S_{t_{1}}=S_{t_{2}}$ if and only if $t_{1}=t_{2}$.

\section{The RNA folding model\label{themodel}}

\subsection{Complex adaptive systems}

The RNA folding model presented below is inspired by basic ideas from the
theory of complex adaptive systems. Specifically, a complex adaptive system
(CAS) is characterized by the presence of a diverse ensemble of components
that engage in local interactions and an autonomous process that selects a
subset of those components for enhancement based on the results of the local
interactions (Levin, 1998). From these component-level dynamics emerge
important global (i.e., system-level) properties such as self-organization
and nonlinearity. Self-organization refers to the emergence of order from
local interactions between the components of a CAS. Self-organization tends
to drive a CAS towards stable configurations or states. In addition, a CAS
exhibits nonlinearity; the rules that govern local interactions between the
components of a CAS change as the CAS evolves (Levin, 1998). Consequently, a
CAS may evolve along any one of a multitude of trajectories and may attain
alternative stable states (Levin, 1998), depending on its particular
evolutionary trajectory. See Levin (1998) and the references therein for
further information on CASs.

\subsection{Details of the model}

The kinetic folding of RNA sequences into secondary structures is viewed
here as a time-series of structural rearrangements (SRs), involving the
formation, dissociation, and \textit{shifting} of individual RNA base pairs.
It is modeled as a hierarchically-structured CAS; at the lowest level of the
hierarchy are RNA bases and base pairs that engage in local stacking
interactions. The results of these stacking interactions determine the
probabilities (or fitnesses) of possible SRs. These probabilities are given
by%
\begin{equation}
P_{f}^{\left( i,j\right) }=e^{-\frac{\Delta G^{ij}}{2RT}},\text{ }
\label{pf}
\end{equation}%
\begin{equation}
P_{d}^{\left( i,j\right) }=\frac{1}{P_{f}^{\left( i,j\right) }},\text{ and}
\label{pd}
\end{equation}%
\begin{equation}
\text{ }P_{s}^{\left( i,j\right) \rightarrow \left( i,k\right)
}=P_{d}^{\left( i,j\right) }P_{f}^{\left( i,k\right) },  \label{ps}
\end{equation}%
where $P_{f}^{\left( i,j\right) }$, $P_{d}^{\left( i,j\right) }$, and $%
P_{s}^{\left( i,j\right) \rightarrow \left( i,k\right) }$ are, respectively,
the probabilities of formation and dissociation of $\left( i,j\right) $, and
of \textit{shifting} $\left( i,j\right) $ into $\left( i,k\right) ,$ $R$\ is
the gas constant, $T$ is the absolute temperature, and $\Delta G^{ij}$ is
the stacking (including single-base stacking) energy associated with $\left(
i,j\right) $. For an isolated base pair $\left( i,j\right) $, 
\begin{equation}
\Delta G^{ij}=\Delta G^{ij}+c\ln \left( j-i\right) ,c\geq 0,  \label{isol}
\end{equation}%
Equation $\left( \text{\ref{isol}}\right) $ takes into account the entropy
of a loop of size $\left( j-i\right) $. A suitable value for the parameter $%
c $ is $1.75$ (Fisher, 1966). Stacking energy calculations are based on the
Turner $3.1$ energy rules (Mathews et al., 1999), at temperature $T=310.5K$,
and on the Turner $2.3$ energy rules (Freier et al., 1986), at other
temperatures. Equations $\left( \text{\ref{pf}}\right) $, $\left( \text{\ref%
{pd}}\right) $, $\left( \text{\ref{ps}}\right) $ are based on the Kawasaki
dynamics (1966); here, the dynamics involve transitions between states
associated with specific local contexts of an RNA secondary structure.

The next level of the hierarchical structure is occupied by SRs. It is at
this level that selection operates. An autonomous stochastic sampling
process (Baker, 1987) periodically (i.e., from one time step to another)
selects a subset of possible SRs for realization based on the fitnesses of
the SRs. The ensemble of possible SRs changes from one time step to another
thereby assuring its diversity (see below). Note that due to the
inter-dependence of local stacking interactions, on which the fitnesses of
SRs depend, it is necessary to ensure that the SRs that are selected for
realization in the same time step be mutually independent. Therefore if
there is an SR involving the base $x_{i}$, then no other SR that involves
the nearest-neighbor bases and base pairs of $x_{i}$ can occur in the same
time step. Furthermore, in order to prevent the formation of pseudoknots SRs
may only involve accessible bases; two bases $x_{i}$ and $x_{j}$, $i<j$, are
accessible if for $k=i+1,\ldots ,j-1$ there is no base pair $\left(
k,l\right) $ $\left( \text{resp}.\left( l,k\right) \right) $ such that $l>j$
(resp. $l<i$).

To understand the model just described, consider the kinetic folding of an
RNA sequence $X$. Denote by $\Re $ the set of possible SRs that are
available for realization in a given time step. $\Re $ will contain as many
elements as there are structures in $N\left( S_{t}\right) $, where $S_{t}$
is the nascent structure of $X$. Each element of $\Re $ is associated with
specific bases and base pairs that belong to that element's "local context".
For instance, an SR that involves the formation of the base pair $\left(
i,j\right) $ is associated with $\left( i,j\right) $ and all
nearest-neighbor bases and base pairs of $\left( i,j\right) $. Stacking
interactions between the bases and base pairs associated with a given SR
determine that SR's probability or fitness. This fitness is used
periodically by an autonomous stochastic process to select a subset of SRs
from $\Re $ for realization.

As SRs are realized (and removed from $\Re $), existing SRs may be become
impractical while new SRs may become possible. For instance, the formation
of $\left( i,j\right) $ in a given time step may make possible the \textit{%
shifting} of $\left( i,j\right) $ into some $\left( i,k\right) $ in the next
time step. Conversely, the dissociation of $\left( i,j\right) $ in a given
time step will render impractical the \textit{shifting} of $\left(
i,j\right) $ into some other base pair in the next time step. New SRs that
become possible are added to $\Re $ while those that become impractical are
removed from $\Re $. This assures the diversity of possible SRs. Note that
the idea that the selective enhancement of components (i.e., SRs in this
case) of a given system leads to their removal (or elimination) as well as
the elimination of other components from that system appears to be at odds
with what takes place in most known CASs. In the present case, the goal of
selection is to, indirectly, enhance the thermal stabilities of the local
contexts associated with SRs.

We note here that the inter-dependence of local stacking interactions, on
which the fitnesses of SRs depend, implies that a given SR may "interact"
with many other SRs. For instance, an SR that involves the base pair $\left(
i,j\right) $ will "interact" with all SRs that involve either of the bases $%
x_{i}$ and $x_{j}$. This implies a relatively high average degree of
"epistatic" interactions between SRs. Results from studies based on random
Boolean networks predict that such a high degree of epistasis leads to
rugged fitness landscapes with numerous attractors (Kauffman and Levin,
1987; Kauffman, 1989). This prediction is consistent with the well-known
rugged nature of RNA folding energy landscapes.

We further note that SRs, as defined in the model, operate at much more
local scales of space than is the case with most existing folding methods
(e.g., see Flamm et al., 2000; Isambert et. al., 2000; Tang et al., 2004).
The fitnesses of SRs depend exclusively on the stacking energies associated
with specific local contexts of the nascent RNA structure, $S_{t}$, and not
on the free-energies of structures found in $N\left( S_{t}\right) $.
Therefore, there is no need for explicit computation of the free-energies of
RNA structures in the present model, in contrast to, e.g., the folding
method of Flamm et al. (2000). We expect the model to reproduce global
characteristics of CASs such as self-organization and nonlinearity. In
particular, RNA molecules are "self-organizing" since, through their own
internal dynamics, they tend to fold into thermodynamically favorable or
stable states. Folding RNA molecules also exhibit nonlinear dynamics, as
evinced by their attainment of alternative stable states. In Section 4, we
will illustrate such nonlinear dynamics using an example based on SV11.

\subsection{Computer implementation of the model}

The above model has been implemented in the computer program \textit{kfold},
which is available from the author upon request. For an input RNA sequence
of length $n$, the program selects $m=1$ SR, if $n\leq 30$, and $m=7$ SRs,
if $n>30$, for realization in each time step. The folding time is
incremented in each time step by the reciprocal of the product of $m$ and
the sum of the fitnesses of all possible SRs. The number of selected SRs $m$
can be adjusted by the user. Note that the choice of $m$ influences the
computer time required to fold an input RNA sequence but has minimal effect
on the qualitative folding kinetics of the sequence (see example in Table
1). Also note that in order to speed up folding simulations, the program
currently only allows the formation of base pairs that can be stacked.
Specifically, a base pair $\left( i,j\right) $ can be stacked if there
exists complementary bases $x_{l}$ and $x_{k},$ $l<k,$ such that either $%
i-k=l-j=1$ or $k-i=j-$ $l=1$.

\begin{equation*}
\begin{tabular}{|l|l|l|l|}
\hline
$m$ & {\small Time steps} & {\small Fraction of A} & {\small Fraction of B}
\\ \hline
$1$ & $249$ & $\allowbreak 0.40\,$ & $0.60$ \\ \hline
$3$ & $89$ & $\allowbreak 0.41\,$ & $0.59$ \\ \hline
$5$ & $134$ & $\allowbreak 0.38\,$ & $0.62$ \\ \hline
$7$ & $169$ & $\allowbreak 0.42$ & $0.58$ \\ \hline
$9$ & $\allowbreak \allowbreak 241$ & $\allowbreak 0.45\,$ & $0.55$ \\ \hline
$11$ & $>30000$ & $0.40$ & $0.60$ \\ \hline
\end{tabular}%
\end{equation*}

\begin{quotation}
Table 1. Influence of $m$ on folding kinetics for the sequence 
\begin{equation*}
GUCCUUGCGUGAGGACAGCCCUUAUGUGAGGGC,
\end{equation*}%
\textit{\ with} $n=33$. It was folded with ((((((((((((((.....))))))))))))))
(A) and ((((((....)))))).((((((....)))))) (B) serving as target structures.
The fraction of simulations that found either structure within the allowed
time scale (i.e., $4\times 10^{4}$ time steps) is similar for different
values of $m$. On the other hand, the number of time steps, which reflects
the amount of computer time required for folding, decreases as $m$ increases
from $1$ to $3$, and subsequently increases with $m$. $50\%$ of simulations
failed to find either target structure for $m=11$. The number of time steps
given for $m=11$ thus represents a lower bound of its actual value. Note
that each data point was averaged from just $500$ folding simulations run at 
$T=310.5K$. Therefore, there may be errors in the data resulting from
limited sampling of possible folding trajectories.
\end{quotation}

\section{Example applications\label{applics}}

We now use the new folding model, as implemented in \textit{kfold}, to study
the effects of base modifications and temperature on the folding kinetics of
the yeast tRNA$^{Phe}$. We also estimate the population dynamics of two
alternative stable states of the synthetic SV11. These examples will
demonstrate that the folding model qualitatively reproduces characteristic
RNA folding dynamics. Note that in the following examples, the folding times
(i.e., mean first passage times) were scaled using experimentally measured
folding times (in $\mu s$) of the hairpin $AAAAAACCCCCCUUUUUU$ (Poerschke,
1974b). This was done in order to allow direct comparisons with folding
times reported in Flamm (1998). Unless otherwise noted, all folding
simulations were run at $T=310.5K$.

\subsection{Influence of base modifications on folding kinetics}

A number of tRNA sequences are known to contain base modifications. Such
modifications, believed to be the consequence of evolutionary optimization,
have been shown to improve the foldabilities of some tRNAs (Flamm, 1998;
Flamm et al., 2000). We have introduced several base modifications to the
individual hairpins of the yeast tRNA$^{Phe}$ sequence and studied the
folding kinetics of the modified sequences. All modified bases were
prohibited from engaging in bond formation and stacking interactions. The
modified sequences are shown in Table 2.

\begin{quotation}
\begin{equation*}
\begin{tabular}{|l|l|l|}
\hline
{\small Sequence} & {\small Modified Hairpins} & {\small Modified Sequence
Positions} \\ \hline
${\small seq1}$ & $1,$ $2$ $\&$ $3$ & $15,17,19,37,38,55,56$ $\&$ $59$ \\ 
\hline
${\small seq2}$ & $1$ $\&$ $2$ & $15,17,19,37$ $\&$ $38$ \\ \hline
${\small seq3}$ & $1$ $\&$ $3$ & $15,17,19,55,56$ $\&$ $59$ \\ \hline
${\small seq4}$ & $2$ $\&$ $3$ & $37,38,55,56$ $\&$ $59$ \\ \hline
${\small seq5}$ & $None$ & $None$ \\ \hline
\end{tabular}%
\end{equation*}

Table 2. Modified tRNA sequences used in this example. Note that hairpins
are labeled from left to right, with "1" representing the left-most hairpin.
\end{quotation}

We found base modifications to elicit substantial improvements in tRNA
foldabilities, in the form of drastic decreases in folding times (see Figure
1). For the unmodified sequence, $seq5$, the fraction of folded sequences
(i.e., sequences that have found the native cloverleaf structure) increased
relatively slowly, reaching about $45\%$ after the first $600us$, and $100\%$
within $4500\mu s$. For all modified sequences, on the other hand, the
fraction of folded sequences increased rapidly, reaching $100\%$ within $%
1500\mu s$. Among these sequences, $seq1$ was the fastest folder with an
estimated folding time of $300\mu s$, while $seq4$ was the slowest folder
with a folding time of $1500\mu s$. The folding times of $seq2$ and $seq3$
were approximately equal (i.e., about $800\mu s$). These observed effects of
base modifications on tRNA folding kinetics are consistent with experimental
data, as well as with predictions made by \textit{kinfold} (Flamm, 1998).
Note that we were able to simulate much longer folding times for tRNA$^{Phe}$%
, up to $4500\mu s$, than was done in Flamm (1998).

\begin{figure}[tbp]
\begin{center}
\includegraphics[width=4.99in]{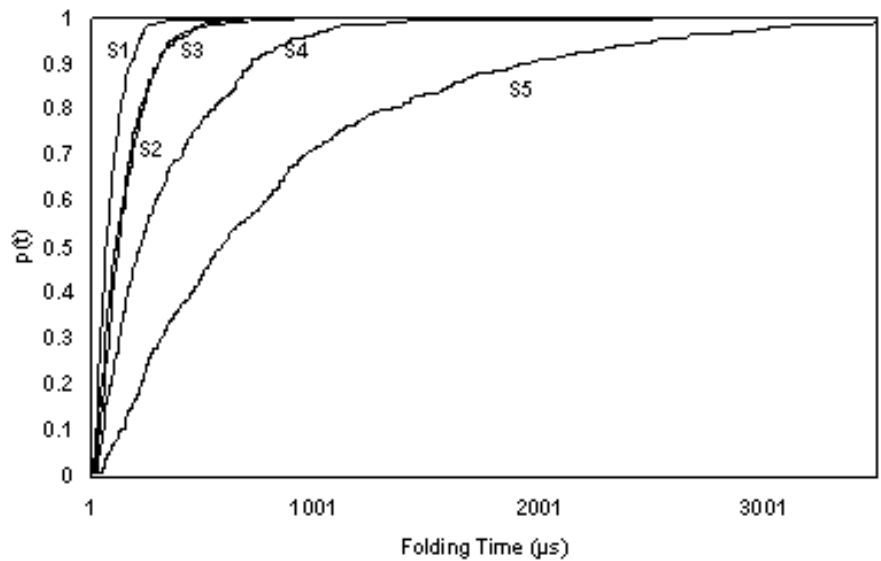}
\end{center}
\caption{Folding kinetics of modified tRNA sequences (see Table 2). For each
sequence, $1000$ \textit{kfold} simulations were run for $1\times 10^{4}%
\protect\mu s$ or until the native cloverleaf was found. $p\left( t\right) $
denotes the fraction of simulations that found the cloverleaf within $t%
\protect\mu s$. Note that the displayed folding times may not be
biologically realistic as they were scaled using the folding kinetics of a
hairpin. Further calibration with experimental data is necessary in order to
ensure biological relevance of the displayed time scales.}
\end{figure}

\subsection{Temperature dependence of folding kinetics}

We have used the folding model to study the temperature dependence of the
folding time $\tau _{f}$ of the yeast tRNA$^{Phe}$. We found a $V$-shaped
temperature dependence of $\tau _{f}$ (see Figure 2), suggesting the
existence of an optimal folding temperature $T_{opt}\approx 313K$. A
possible explanation for the existence of $T_{opt}$ is as follows: At
temperatures $T>T_{opt}$, there are numerous structures with similar
free-energies as the native cloverleaf. The native cloverleaf is therefore
relatively unstable at temperatures above $T_{opt}$ and may be associated
with a much smaller basin of attraction in the energy landscape. It
therefore takes the folding tRNA longer to "find" the cloverleaf among the
ensemble of nonnative states. On the other hand, at temperatures $T<T_{opt}$%
, the stability of nonnative structures increases leading to a growth in the
number and, perhaps, sizes of nonnative basins of attraction in the energy
landscape. These nonnative basins of attraction may decrease the folding
tRNA's chances of finding the native state. Both scenarios (i.e., $T>T_{opt}$
and $T<T_{opt}$) lead to suboptimal native state accessibilities. Meanwhile,
the maximal accessibility of the native state that is evident at $T=T_{opt}$
suggests the existence at $T_{opt}$ of optimal balance between the thermal
stabilities of native and nonnative states. Note that $T_{opt}$ is close to
the optimal growth temperature range for many yeast species. Detailed
analysis of the temperature dependence of folding kinetics for tRNAs and
other functional RNAs from various organisms will allows us to determine if
this observation is a consequence of evolutionary optimization or simply a
chance occurrence.

\begin{figure}[tbp]
\begin{center}
\includegraphics[width=4.31in]{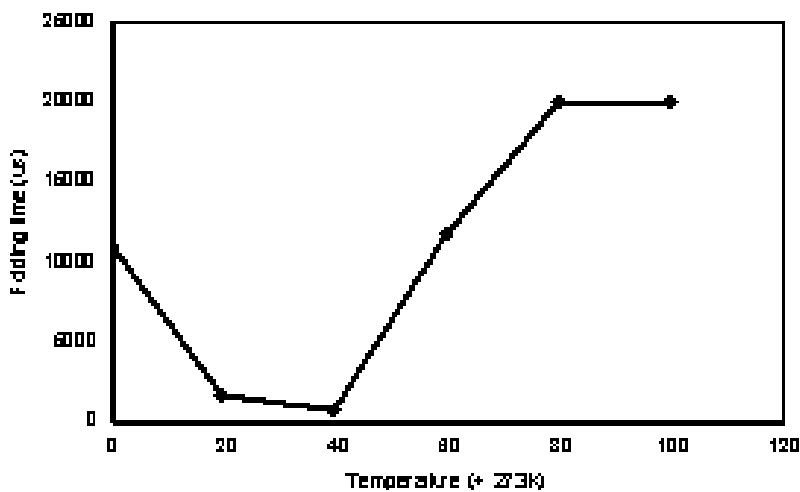}
\end{center}
\caption{Temperature dependence of the folding time for tRNA obtained from
folding simulations run at $T=273,$ $293,$ $313,$ $333,$ $353,$ and $373K$.
None of the simulations found the ground state at $T=353$ and $373K$ within $%
2\times 10^{4}\protect\mu s$. The folding times shown for these temperatures
therefore represent lower-bounds of their actual values. Energy calculations
were based on the Turner $2.3$ energy parameters (Freier et al., 1986), for
which extrapolations to temperatures other than $310.5K$ are readily
available. Each data point was averaged from just 500 folding simulations.
Therefore, there may be errors in the data resulting from limited sampling
of possible folding trajectories.}
\end{figure}

\subsection{(Meta)stable states\label{meta}}

During kinetic folding, some RNA molecules may get trapped in long-lived,
nonnative states called metastable states/conformations. Examples of such
metastable RNA molecules include riboswitches that regulate gene expression
in bacteria by switching between alternative stable conformations
(Vitreshack et al., 2004). By adopting a repressing conformation, a
riboswitch can elicit the premature termination of DNA transcription or the
inhibition of protein translation. Detailed \textit{in silico} analysis of
the folding kinetics of a metastable RNA molecule can provide insight into
its functionality. For instance, Nagel et al. (1999) used computer
simulations to identify a metastable structure of the \textit{Hok} mRNA that
mediates apoptosis in plasmid R1-free cells. The predictions made by their
simulations were in good agreement with experimental data (Nagel, 1999). See
Higgs (2000) for other examples of how computer simulations have yielded
important insight into the folding kinetics of metastable RNA molecules.

SV11 is a 115nt synthetic RNA species that exists in two alternative stable
states, a rod-like stable conformation and a multi-component metastable
conformation (see Figure 3). While the metastable conformation is a template
for Q$\beta $ replicase, the stable conformation is not (Zamora et al.,
1995). Several authors have previously performed \textit{in silico} analysis
of the folding kinetics of SV11. Some of the authors (Flamm, 1998; Flamm et
al., 2000) successfully predicted the existence of the metastable
conformation while others (e.g., see Gultyaev et al., 1990; Morgan \& Higgs,
1996) could only do so if folding was constrained to occur in conjunction
with transcription. However, none of the authors obtained detailed
theoretical estimates of the population dynamics of the molecule's two
alternative stable states.

\begin{figure}[tbp]
\begin{center}
\includegraphics[width=4.51in]{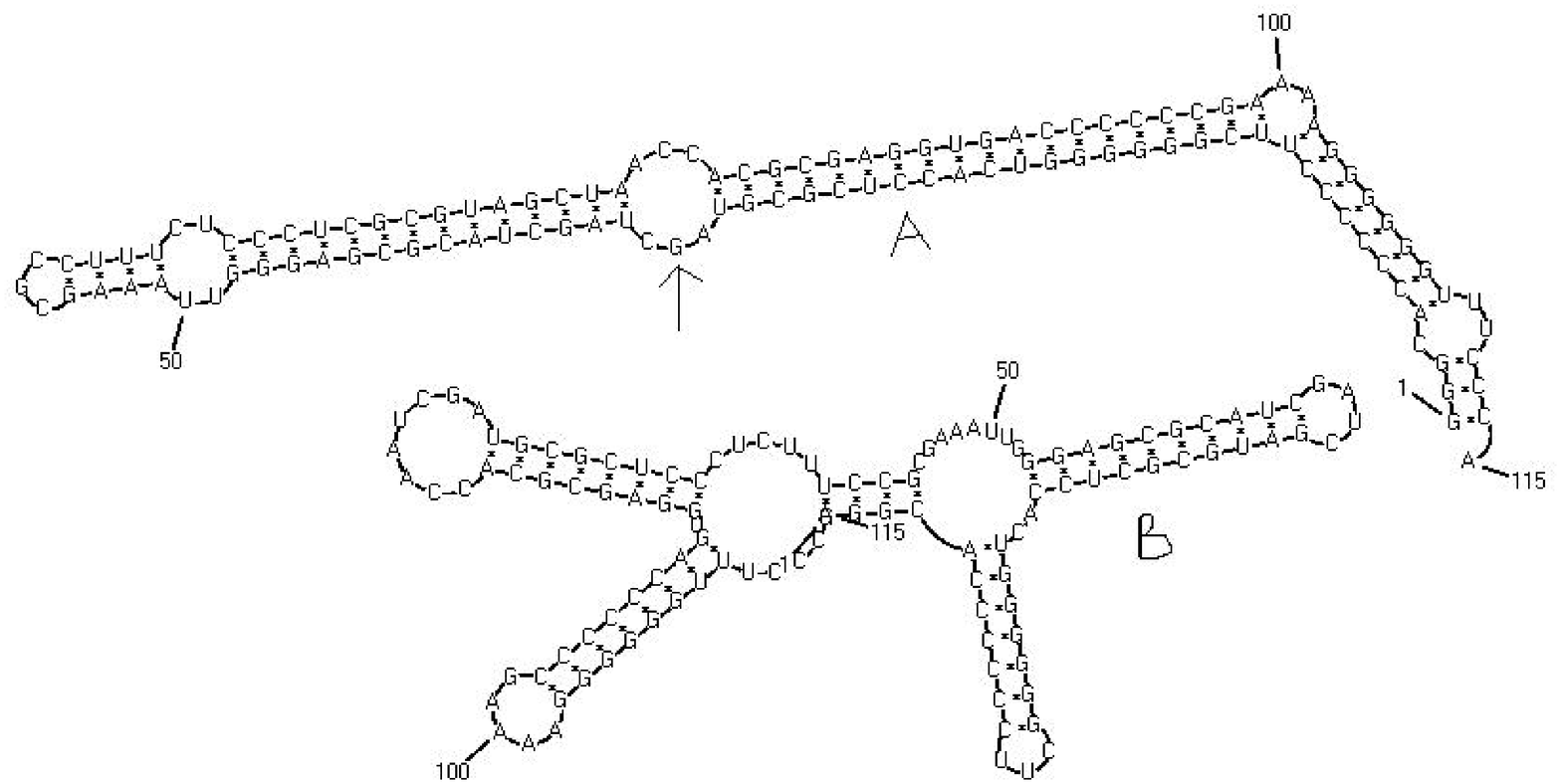}
\end{center}
\caption{The stable (A) and metastable (B) structures of SV11. Note that a
33GC79 base pair (see arrow) was removed from the stable structure since 
\textit{kfold} currently only allows base pairs that can be stacked; there
are no complementary bases $x_{l}$ and $x_{k},$ $l<k,$ in the SV11 sequence
such that either $33-k=l-79=1$ or $k-33=79-$ $l=1$.}
\end{figure}

Here we report detailed estimates of the population dynamics of the stable
and metastable conformations of SV11. As shown in Figure 4 the population of
the metastable state increases steadily, reaching a maximum after about $%
2000\mu s$. In experiments performed using \textit{kinfold}, Flamm (1998)
reported that the fraction of the metastable state reached about $16\%$
after $500\mu s$. This estimate is consistent with the results shown in
Figure 4. However, we were able to fold the molecule for much longer, up to $%
1.5\times 10^{4}\mu s$, than was done in Flamm (1998). This allowed us to
estimate the population dynamics of not just the metastable conformation but
also of the stable native conformation. The ratio of the fraction of
simulations that found the native conformation to the fraction that found
the metastable conformation in the time scale of the simulation was
approximately $3$ \ to $1$. Note that the accuracy of these results can be
tested directly, in the laboratory.

\begin{figure}[tbp]
\begin{center}
\includegraphics[width=4.97in]{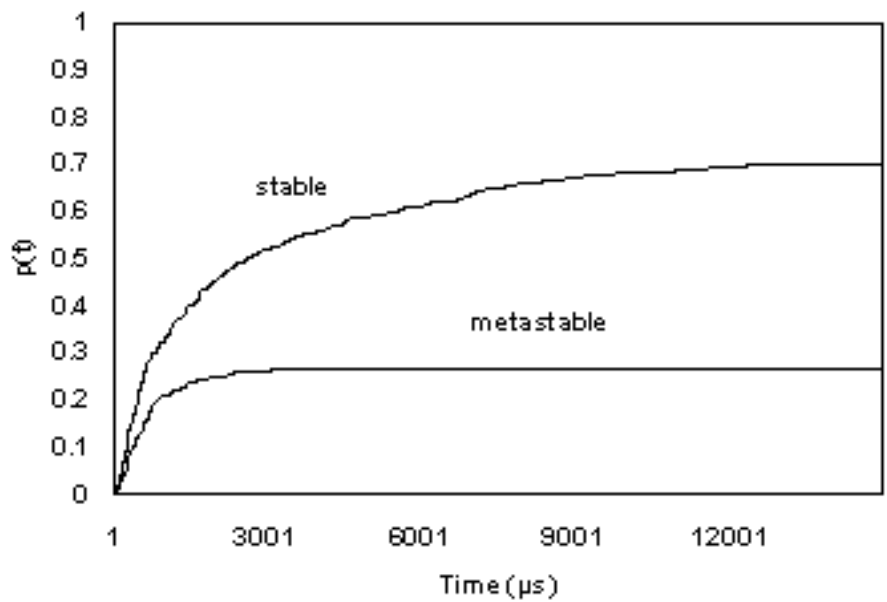}
\end{center}
\caption{Folding kinetics of SV11, obtained from 1000 \textit{kfold}
simulations. Each simulation was stopped after it found either of the
molecule's two alternative stable states. $p\left( t\right) $ denotes the
fraction of simulations that found either state within $t\protect\mu s$.
Note that the displayed folding times may not be biologically realistic as
they were scaled using the folding kinetics of a hairpin. Further
calibration with experimental data is necessary in order to ensure
biological relevance of the displayed time scales.}
\end{figure}

\section{Discussion\label{concl}}

In this paper, we introduced a new model for the kinetic folding of RNA
sequences into secondary structures that was inspired by the theory of
complex adaptive systems. In the folding model, RNA bases and base pairs
engage in local stacking interactions that determine the probabilities (or
fitnesses) of possible RNA structural rearrangements (SRs). Meanwhile,
selection operates at the level of SRs; an autonomous stochastic sampling
process periodically selects a subset of possible SRs for realization based
on the fitnesses of the SRs. Several examples were used to illustrate the
applicability of the model. In particular, certain base modifications were
shown to substantially improve the foldability of tRNA$^{Phe}$. In addition,
a characteristic optimal folding temperature $T_{opt}\left( \approx
313K\right) $ of tRNA$^{Phe}$ was identified. Furthermore, the model was
used to confirm previous experimental results (Zamora et al., 1995)
regarding the existence of two alternative stable states of the Q$\beta $
variant SV11, and to obtain (experimentally verifiable) estimates of the
population dynamics of those states. The above examples demonstrated, among
other things, the emergence from (local) SRs of nonlinear RNA folding
dynamics (i.e., the realization of alternative stable states). Other
possible applications of the model are discussed below.

The analysis of properties of fitness landscapes is currently of interest to
researchers in a wide range of fields including the life, computer, and
social sciences (e.g., see Hadany \& Beker, 2003; McCarthy, 2004; Skellett
et al., 2005). A number of interesting general features of these landscapes
such as positive correlations between the degree of epistasis, the number of
local optima, and the expected value of the global optimum have thus far
been elucidated in the context of adaptive walks on landscapes generated by
random Boolean networks (RBNs) (Kauffman \& Levin, 1987; Kauffman, 1989;
Skellett et al., 2005) . The simplicity of RBNs and their accessibility to
some degree of mathematical analysis make them convenient for use as
generators of fitness landscapes. However, some of the general properties of
such RBN-generated landscapes may differ from those of landscapes found in
Nature. RNA folding kinetics, as modeled in this paper, could serve as a
generator of, and therefore assist the analysis of properties of "natural"
fitness landscapes (i.e., RNA folding energy landscapes). Note that this
proposed use of RNA in the investigation of fitness landscapes differs from
the previous related use of the RNA sequence to structure mapping (Schuster
\& Stadler, 1994), which was not based on folding kinetics.

The model could also be used to study the RNA sequence to structure (or
genotype to phenotype) mapping, from a kinetics perspective. Previous
investigations, based on the thermodynamics of RNA folding, have made
several important findings about the RNA sequence to structure mapping such
as the existence of (1) extended neutral networks of sequences that fold
into the same secondary structures (2) few common or "typical" structures,
realized with relatively high frequencies, and (3) many rare structures that
have little or no evolutionary significance (Schuster et al., 1994; Schuster
et al., 1998). These findings, together with results from
thermodynamics-based RNA optimization experiments (Fontana \& Schuster,
1998), have confirmed previous hypotheses on, and shed light into several
important features of the process of molecular evolution, including the role
of neutrality in adaptation and the existence of continuous/discontinuous
transitions or punctuated equilibria in evolutionary trajectories. It would
be interesting to determine how the nature of the RNA sequence to structure
mapping, as obtained from thermodynamic folding experiments, changes when
RNA folding kinetics is taken into account. It is also possible that an
investigation of the genotype to phenotype mapping based on the kinetics of
RNA folding will yield further insight into the process of molecular
evolution.

\section{Acknowledgements}

This work was funded in part by DOE-ER63580 and by RCMI Grant no. RR017581.
The author thanks Dr. Asamoah Nkwanta and anonymous reviewers for their
useful comments on an earlier version of the manuscript.

\section{References}

\begin{description}
\item Abrahams, J.P., van den Berg, M., van Batenburg, E., Pleij, C, 1990.
Prediction of RNA secondary structure, including pseudoknotting, by computer
simulation.\ Nucl. Acids Res. 18, 3035-3044.

\item Baker, J. E., 1987. Reducing bias and inefficiency in the selection
algorithm. In: Grefenstette, E. (Ed.), Proceedings of the Second
International Conference on Genetic Algorithms and their Application.
Lawrence Erlbaum Associates, Hillsdale, pp. 14-21.

\item Cannone, J.J., Subramanian, S., Schnare, M.N., Collett, J.R., D'Souza,
L.M., Du, Y., Feng, B., Lin, N., Madabusi, L.V., Muller, K.M., Pande, N.,
Shang, Z., Yu, N., and Gutell, R.R., 2002. The comparative RNA web (CRW)
site: an online database of comparative sequence and structure information
for ribosomal, intron, and other RNAs. BMC Bioinformatics 3, 2.

\item Fisher, M.E., 1966. Effect of excluded volume on phase transitions in
biopolymers. J. Chem. Phys. 45, 1469--1473.

\item Flamm, C., 1998. Kinetic folding of RNA. Doctoral dissertation.
University of Vienna, Austria.

\item Flamm, C., Fontana, W., Hofacker, I.L., Schuster, P., 2000. RNA
folding at elementary step resolution. RNA 6, 325-338.

\item Fontana, W., Schuster, P., 1998. Continuity in evolution: On the
nature of transitions. Science 280, 1451-1455.

\item Freier, S.M., Kierzek, R., Jaeger, J.A., Sugimoto, N., Caruthers,
M.H., Neilson, T., Turner, D.H. 1986. Improved parameters for predictions of
RNA RNA duplex stability. Proc. Natl. Acad. Sci. 83, 9373-9377.

\item Gratias, A., Betermier, M., 2003. Processing of double-strand breaks
is involved in the precise excision of paramecium internal eliminated
sequences. Mol. Cell. Biol. 7152--7162.

\item Gultyaev, A.P., van Batenburg, F.H.D., Pleij, C.W.A., 1995. The
influence of a metastable structure in plasmid primer RNA on antisense RNA
binding kinetics. Nucl. Acids Res. 23, 3718-3725.

\item Gultyaev, A.P., van Batenburg, F.H.D., Pleij, C.W.A., 1990. The
computer simulation of RNA folding pathways using a genetic algorithm. J.
Mol. Biol., 250, 37--51.

\item Hadany, L., Beker, T., 2003. Fitness-associated recombination on
rugged adaptive landscapes. J. Evol. Biol. 16, 862-870.

\item Higgs, P.G., 2000. RNA secondary structure: physical and computational
aspects. Quat. Rev. Biophys. 33, 199-253.

\item Hofacker, I.L, Fontana, W., Stadler, P.F., Bonhoffer, S., Tacker, M.,
Schuster, P., 1994. Fast folding and comparison of RNA secondary
structures.\ Monatsch. Chem. 125, 167-188.

\item Isambert, H., Siggia, E.D., 2000. Modeling RNA folding paths with
pseudoknots: application to hepatitis delta virus ribozyme. Proc. Natl.
Acad. Sci. 97, 6515-6520.

\item Kawasaki, K., 1966. Diffusion constants near the critical point for
time-dependent Ising models. Phys. Rev. 145, 224-230.

\item Kauffman, S.A., 1989. Adaptation on rugged landscapes. In D. Stein
(ed.) Lectures in the Sciences of Complexity, lecture volume 1, pp. 527-618.
Addison-Wesley, Redwood City.

\item Kauffman, S.A., Levin, S.A., 1987. Towards a general theory of
adaptive walks on rugged landscapes. J. Theor. Biol. 128, 11-45.

\item Lee, N.S., Dohjima, T., Bauer, G., Li, H., Li, M.J., Ehsani, A.,
Salvaterra, P., Rossi, J., 2002. Expression of small interfering RNAs
targeted against HIV-1 rev transcripts in human cells. Nat. Biotechnol. 20,
500-505.

\item Levin, S.A., 1998. Ecosystems and the biosphere as complex adaptive
systems. Ecosystems 1, 431-436.

\item Mathews, D.H, Sabina, J., Zuker, M., Turner, D.H., 1999. Expanded
sequence dependence of thermodynamic parameters provides robust prediction
of RNA secondary structure.\ J. Mol. Biol. 288, 911-940.

\item Mathews, D.H., Disney, M.D., Childs, J.L., Schroeder, S.J., Zuker, M.,
Turner, D.H., 2004. Incorporating chemical modification constraints into a
dynamic programming algorithm for prediction of RNA secondary structure.
Proc. Natl. Acad. Sci. 101,7287-7292.

\item McCarthy, I.P., 2004. Manufacturing strategy: understanding the
fitness landscape. Intl. J. Op. Prod. Mgt. 24, 124-150.

\item Metropolis, N., Rosenbluth, A. W., Rosenbluth, M. N., Teller, A. H.,
Teller, E., 1953. Equation of state calculations by fast computing machines.
J. Chem. Phys. 21, 1087--1092.

\item Mironov, A., Kister, A., 1985. A kinetic approach to the prediction of
RNA secondary structures. J. Biomol. Struct. Dyn. 2, 953--962.

\item Mochizuki, K., Fine N.A., Fujisawa T., Gorovsky, M.A., 2002. Analysis
of a piwi-related gene implicates small RNAs in genome rearrangement in
Tetrahymena. Cell 110, 689-699.

\item Morgan, S.R., Higgs, P.G., 1996. Evidence for kinetic effects in the
folding of large RNA molecules. J. Chem. Phys. 105, 7152--7157.

\item Nagel, J.H.A., Gultyaev, A.P., Gerdes, K., Pleij, C.W.A., 1999.
Metastable structures and refolding kinetics in hok mRNA of plasmid R1. RNA
5, 1408--1419.

\item Ndifon, W., Nkwanta, A., 2005. An agent-oriented simulation of RNA
folding and its application to the analysis of RNA conformational spaces.
In: Yilmaz, L. (Ed.), Proceedings of the Agent-Directed Simulation Symposium
of the 2005 Spring Simulation Multiconference, SCS Press, San Diego, pp.
198-204.

\item Poerschke, D., 1974a. Model calculations on the kinetics of
oligonucleotide double helix coil transitions. Evidence for a fast chain
sliding reaction. Biophys. Chem. 2, 83--96.

\item Poerschke, D., 1974b. Thermodynamic and kinetic parameters of an
oligonucleotide hairpin helix. Biophys. Chem. 1, 381--386.

\item Schuster, P., Fontana, W., Stadler, P., Hofacker, I.L., 1994. From
sequences to shapes and back: A case study in RNA secondary structures.
Proc. Roy. Soc. (London) B 255, 279-284.

\item Schuster, P., Fontana, W., 1998. Chance and necessity: Lessons from
RNA. Physica D 133, 427-452.

\item Schuster, P., Stadler, P., 1994. Landscapes: complex optimization
problems and biopolymer structure. Comput. Chem. 18, 295-314.

\item Skellett, B., Cairns, B., Geard, N., Tonkes, B., Wiles, J., 2005.
Maximally rugged NK landscapes contain the highest peaks. In: Beyer, H.G.,
O'Reilley, U.M., Arnold, D.V., et al. (Eds.), Proceedings of the 2005
Genetic and Evolutionary Computation Conference, ACM Press, New York, pp.
579-584.

\item Tang, X., Kirkpatrick, B., Thomas, S., Song, G., Amato, N., 2004.
Using motion planning to study RNA folding kinetics. In Proceedings of the
International Conference on Computational Molecular Biology (RECOMB). ACM
Press, San Diego, pp. 252-261.

\item Vitreschack, A.G., Rodinov, D.A., Mironov, A.A., Gelfand, M.S., 2004.
Riboswitches: the oldest mechanism for the regulation of gene expression?
Trends Genet. 20, 44-50.

\item Woese, C.R., Pace, N.R., 1993. Probing RNA structure, function, and
history by comparative analysis. In: Gesteland, R.F., Atkins, J.F. (Eds.),
The RNA World. Cold Spring Harbor Laboratory Press, New York, pp. 91-117.

\item Wolfinger, M.T., Svrcek-Seilera, W.A., Flamm, C., Hofacker, I.L.,
Stadler, P.F., 2003. Efficient computation of RNA folding dynamics. J. Phys.
A 37, 4731-4741.

\item Xayaphoummine, A., Bucher, T., Thalmann, F., Isambert, H., 2003.
Prediction and statistics of pseudoknots in RNA structures using exactly
clustered stochastic simulations. Proc. Natl. Acad. Sci. 100, 15310--15315.

\item Yang, G., Thompson, J.A., Fang, B., Liu, J., 2003. Silencing of H-ras
gene expression by retrovirus-mediated siRNA decreases transformation
effciency and tumor growth in a model of human ovarian cancer. Oncogene 22,
5694-5701.

\item Zamora, H., Luce, R., Biebricher, C.K., 1995. Design of artificial
short-chained RNA species that are replicated by Q replicase. Biochemistry
34, 1261--1266.

\item Zhang, W., Chen, S.J., 2002. RNA hairpin-folding kinetics. Proc. Natl.
Acad. Sci. 99, 1931--1936.
\end{description}

\end{document}